\documentclass[RNAAS]{aastex631}
\usepackage{lineno}

\begin{document}

\title{Finding Passive Galaxies in HI-MaNGA: The Impact of Star-Formation Rate Indicator}

\author{Nora Salem}
\affiliation{Departments of Physics and Astronomy, Haverford College,
370 Lancaster Ave., Haverford, PA 19041, USA}

\author[0000-0003-0846-9578]{Karen L. Masters}
\affiliation{Departments of Physics and Astronomy, Haverford College,
370 Lancaster Ave., Haverford, PA 19041, USA}
\email{klmasters@haverford.edu}

\author{David V. Stark}
\affiliation{Space Telescope Science Institute, 3700 San Martin Dr Baltimore, MD, 21218, USA}

\author[0000-0002-2343-0682]{Anubhav Sharma}
\affiliation{Departments of Physics and Astronomy, Haverford College,
370 Lancaster Ave., Haverford, PA 19041, USA}

\begin{abstract}

HI-rich galaxies typically have high star-formation rates (SFR), but there exist interesting HI-rich and low star-forming (low-SF) galaxies. Previous work on a sample of these galaxies identified from HI-MaNGA (HI follow-up to the MaNGA survey) using an infrared indicator of specific-SFR (sSFR; namely W2$-$W3~$<2$) could find no single physical process to explain their unusual behaviour. The method by which galaxies are identified as low sSFR may be important in this conclusion. In this Research Note, we explore how an HI-rich, low sSFR sample of HI-MaNGA galaxies differs using H$\alpha$, single stellar population, and ultraviolet estimators of SFR. We find that samples are statistically similar to each other so long as W2$-$W3~$<2$ is interpreted as corresponding to sSFR$<10^{-11.15}$ yr$^{-1}$.

\end{abstract}

\section{Introduction}
It is well known that specific star formation rates (sSFR = SFR/$M_\star$, where $M_\star$ is stellar mass) in galaxies are positively correlated with their cold gas (neutral hydrogen, HI) fraction \citep[$M_{\rm HI}/M_\star$, e.g.][]{Doyle2006,Zhou2018,Saintonge2022}. However, there exist galaxies that are HI-rich but not forming stars at the expected rate \citep[e.g.][]{Parkash2019,Sharma2023,Li2024}. These provide interesting samples to understand the gas-to-star pipeline, and mechanisms which prevent star formation in galaxies. Previous work has found they have high low-ionization emission-line regions (LIER) fractions \citep[][]{Parkash2019}, or are preferentially located at the center of low-mass halos \citep{Li2024}. The sample in \citet{Sharma2023}, based on the Mapping Nearby Galaxies at Apache Point Observatory \citep[MaNGA, ][]{Bundy2015} data, showed a mixture of evidence for recent accretion of gas, enhanced active galactic nucleus (AGN) feedback (via LIER and/or red geyser fractions), bar quenching, and HI preferentially found in high angular momentum, large-radius, low-density disks. To date there is no single defining physical property correlated with quiescent SF behavior in otherwise gas-rich galaxies.

Details of sample selection are crucial to understand a rare subset of galaxies. In this Research Note, we explore HI-rich galaxies in MaNGA \citep{Sharma2023}, identifying low-SF galaxies based on infrared (IR) as well as H$\alpha$ and UV emission, and single stellar population (SSP) models from Pipe3D \citep{Sanchez2016,Sanchez2022}. We seek to understand how the results of \citet{Sharma2023}, who used IR to identify low-SF galaxies,  might change using alternative SF estimates.

\section{Data and Sample Selection} 
We follow \citet{Sharma2023}, creating an HI-rich ($\log({M_{HI}/M_{\odot}}) > 9.3 $) sample using HI-MaNGA; an HI follow-up to the MaNGA survey \citep{Masters2019,Stark2021}. We use a version of HI-MaNGA, with slightly more galaxies than DR3\footnote{Version DR3.1 found at {\tt https://greenbankobservatory.org/science/gbt-surveys/hi-manga/}}, and obtain a sample of 2232 HI-rich MaNGA galaxies.

\begin{figure}
\begin{center}
    \includegraphics[trim={0mm 0 1cm 1cm},clip,scale=.5]{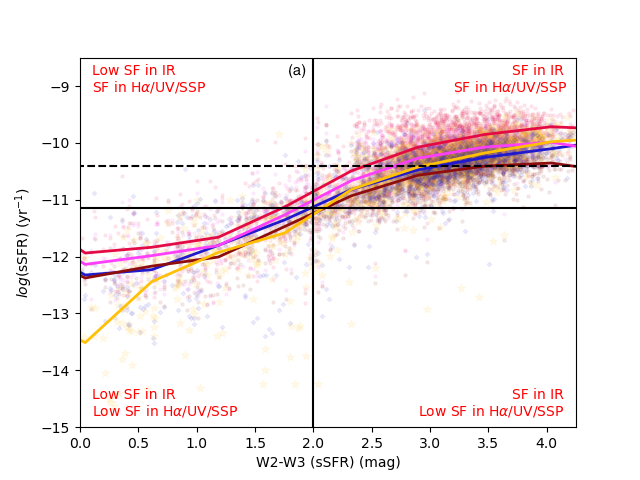}
    \includegraphics[trim={0cm 0 1cm 1cm},clip,scale=.5]{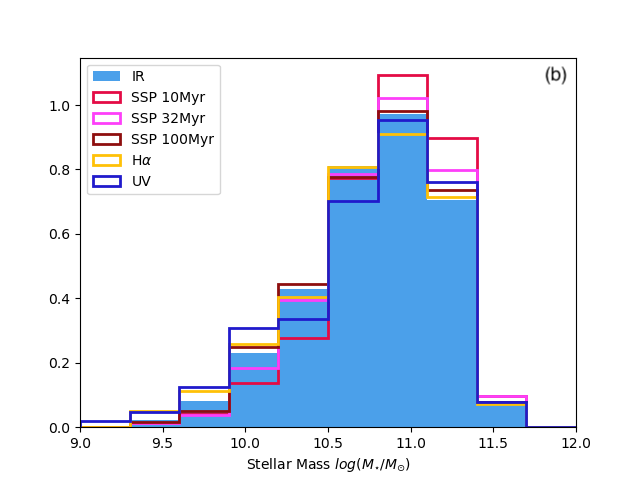}
    \includegraphics[trim={0cm 0 1cm 1cm},clip,scale=.5]{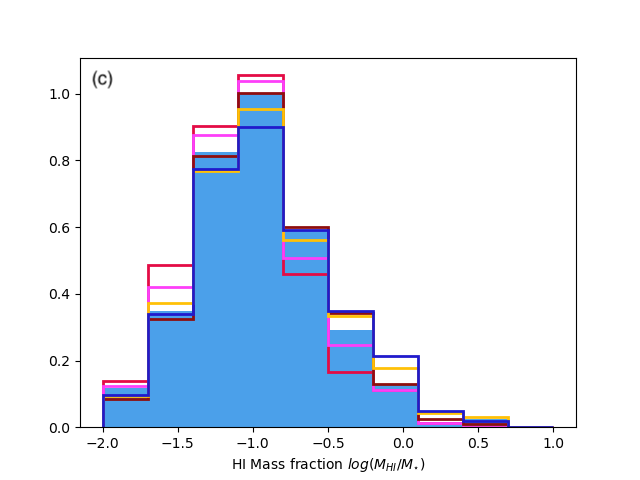}
    \includegraphics[trim={0cm 0 1cm 1cm},clip,scale=.5]{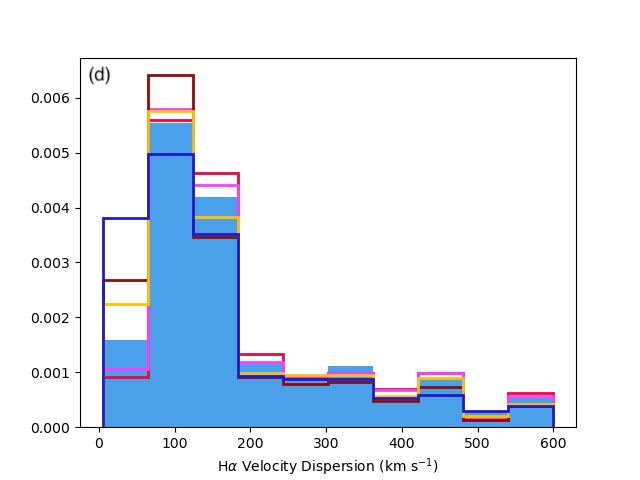}
    \caption{(a): Scatter plot of H$\alpha$-, SSP-, and UV-derived sSFRs against W2-W3 (IR) for HI-rich MaNGA galaxies (colours as in the legend in panel (b)). Overplotted are lines of binned averages. The solid vertical line shows W2-W3=2, and the horizontal lines shows sSFR=$10^{-11.15}$ (solid) or $10^{-10.4}$ (dashed). (b): Histograms of stellar masses for all HI-rich low-SF samples. (c/d): As (b) for  HI mass fraction and H$\alpha$ velocity dispersion. \label{figure}}
\end{center}
\end{figure}

We obtain integrated SFR values from the MaNGA H$\alpha$ flux and Pipe3D SSP models (at times $t < $ 100 Myr, 32 Myr, and 10 Myr, or SSP100, SSP32, and SSP10 respectively), and from UV and IR photometry. We use the DR17 MaNGA Data Analysis Pipeline \citep[DAP,][]{Westfall2019}, and Pipe3D analysis \citep[v3\_1\_1,][]{Sanchez2016,Sanchez2022}, matching with the Wide-field Infrared Survey Explorer AllWISE Data Release \citep[WISE,][]{Wright2010} and GALEX-SDSS-WISE Legacy Catalog-1 \citep[GSWLC-1,][]{Salim2016,Salim2018}. Panel (a) of Figure \ref{figure} shows the H$\alpha$, SSP, and UV estimated sSFRs against W2-W3 (IR color). We observe a broad positive correlation with significant scatter. Galaxies with an IR color of W2-W3 $<$ 2.0 have previously been linked to sSFR~$< 10^{-10.4} {\rm yr}^{-1}$ \citep{Parkash2019,Sharma2023}, based on calibrations by \citet{Cluver2014,Brown2017}.  However when we compare W2-W3 with other sSFR measures available to us from MaNGA, we find that W2-W3~$<2$ in our sample better corresponds to a threshold of sSFR~ $< 10^{-11.15}~ {\rm yr}^{-1}$ (the average values across W2$-$W3$=2\pm0.2$ is sSFR$=10^{-11.15}~{\rm yr}^{-1}$, with a range $10^{-(10.9-11.4)} ~{\rm yr}^{-1}$ for the different indicators). Using this limit we obtain samples of HI-rich but low SF galaxies with $N=$~342, 243, 273, 392, 299, and 288 for SF based on H$\alpha$, SSP10, SSP32, SSP100, UV, and IR respectively. There are still a small number of galaxies that are indicated to be SF via one indicator, but low SF according to a different indicator:  between 1\%-4\% of the H$\alpha$, SSP, and UV SF galaxies are found to be quiescent in IR, while 1\%-5\% of them are found to be SF in IR.

\section{Comparison of HI-Rich Low-SF Galaxy Samples}
We conduct Kolmogorov-Smirnov (K-S) tests\footnote{SciPy 2-sample K-S function} comparing properties of galaxies in the H$\alpha$-, SSP-, and UV-based low-SF samples against the IR-based sample. We consider many of the same global galaxy properties presented in \citet{Sharma2023}: the H$\alpha$- and stellar-derived rotational velocity, velocity dispersion, and absolute velocity asymmetry, as well as stellar mass and HI mass fraction. Figure \ref{figure} shows histograms for three of these. We find that 38/40 K-S tests show distributions which are statistically indistinguishable. Only the H$\alpha$ velocity dispersion in samples constructed using the SSP100 and UV indicators are statistically different from the IR-based sample ($p=0.001$ and $p=0.002$ respectively; see lower right panel of Figure \ref{figure}). These samples both have H$\alpha$ velocity dispersion distributions which are skewed to smaller values.

\section{Discussion and Conclusion}
Using a sample of HI-rich galaxies from HI-MaNGA, we consider how low-SF samples constructed using six different indicators of sSFR (based on optical, UV and IR data) compare. 

We find the W2-W3$<2$ threshold used by both \citet{Parkash2019,Sharma2023} to identify HI-rich low-SFR galaxies best corresponds to sSFR~$< 10^{-11.15} {\rm yr}^{-1}$ for our other measures of sSFR. This is 0.75 dex smaller than previously suggested \citep[][based on \citealt{Cluver2014,Brown2017}]{Parkash2019}. Most of the SFR indicators used here (H$\alpha$, SSP10, SSP32, SSP100) are based on MaNGA data, and measured only in the MaNGA bundle, which extends to only 1.5$R_e$ or 2.5$R_e$ \citep{Wake2017} and may miss outer SF, however we do also measure a similar offset using wider area UV photometry from GALEX \citep{Salim2016}. 

We find some scatter in low sSFR galaxies identified with the different methods, however 95-99\% of galaxies are in common, and across 40 different K-S tests (comparing six samples across eight different global galaxy properties to the IR based sample) we find only two revealing statistically distinguishable galaxy properties (both H$\alpha$ velocity dispersion). We conclude that the choice of \citet{Sharma2023} to use IR to determine low star-formation rate is not likely to produce statistically different results in other indicators.

\end{document}